\documentclass[pdflatex,sn-mathphys-num]{sn-jnl}

\usepackage{graphicx}
\usepackage{multirow}
\usepackage{amsmath,amssymb,amsfonts}
\usepackage{amsthm}
\usepackage{mathrsfs}
\usepackage[title]{appendix}
\usepackage{xcolor}
\usepackage{textcomp}
\usepackage{manyfoot}
\usepackage{booktabs}
\usepackage{algorithm}
\usepackage{algorithmicx}
\usepackage{algpseudocode}
\usepackage{listings}

\usepackage[strings]{underscore}

\theoremstyle{thmstyleone}
\newtheorem{theorem}{Theorem}

\theoremstyle{thmstyletwo}

\theoremstyle{thmstylethree}

\begin{document}

\title[Article Title]{Lieb's theorem for Bose Hubbard models}

\author*[1]{\fnm{Zhong-Chao} \sur{Wei}}\email{zcwei@thp.uni-koeln.de}

\author[2]{\fnm{Chong} \sur{Zhao}}\email{chong.zhao@sdu.edu.cn}

\affil*[1]{\orgname{Qingdao Binhai Univeristy}, \orgaddress{\city{Qingdao}, \postcode{266555}, \country{China}}}

\affil[2]{\orgdiv{School of Mathematics}, \orgname{Shandong University}, \orgaddress{\city{Jinan}, \postcode{250100}, \country{China}}}

\abstract{Using a cone-theoretical method, we prove the uniqueness of the ground
state for two Bose Hubbard models. The first model is the usual Bose
Hubbard model with real hopping coefficients and attractive interactions.
The second model is a two-component Bose Hubbard model. Under certain conditions,
we show that the ground state in the subspace with particle number $N=2n$($n$
is a positive integer) is unique for both models. For the second model,
we show that the ground state has spin along the z-axis $S^{z}=0$.
When the hopping coefficients are real, it has zero spin quantum number,
i.e., it is a singlet. Our proofs work equally well for any arbitrary finite-size lattice.}

\keywords{Lieb's theorem, Bose Hubbard model, convex cone, Perron--Frobenius theorem}

\maketitle

\section{Introduction}

Understanding the ground state property of interacting quantum lattice
models remains a great challenge in fundamental research of physics.
Analytical results are rare and valuable. Using a Perron--Frobenius
type of arguments, the uniqueness of the ground states has been proved
in the past a few decades for Heisenberg model\cite{lieb_ordering_1962},
fermionic Hubbard model\cite{lieb_two_1989}, and an interacting spinless
fermion model\cite{wei_ground_2015}. For the last two models, the
proofs exploited an important property of operators and quantum states
called reflection positivity\cite{kondratiev_reflection_2006,biskup_reflection_2009,jaffe_reflection_2015,jaffe_characterization_2016}.

In this work, using a cone-theoretical method, we generalize the famous
method that was first proposed by Elliott H. Lieb\cite{lieb_two_1989}. Our method is based
on the Perron--Frobenius theorem for convex cones\cite{tam_matrices_2006}.
We apply our method to two Bose Hubbard models and show the uniqueness
of their ground states. We point out that the uniqueness of the ground
states of the two models is closely related to two different positivities
of the ground states, the Fock space positivity and the Fock space
reflection positivity, respectively.

The theory of convex cones has a wide range of applications in several fields,
including structural engineering, control systems, signal processing,
and optimization. Convex cones are used to model the strength of materials
in structural engineering. Convex cones are also used to represent the set
of feasible solutions to various control and signal processing problems.
By using convex cones, mathematicians can find solutions to optimization problems
quickly and efficiently.

The two Bose Hubbard models studied in this work are closely related
to the Bose-Einstein condensation in ultracold atomic gases. For the
second model, namely the two-component Bose Hubbard model, the system
in thermodynamic limits usually undergoes a phase transition when the
parameters change. Therefore, studying this model is also very useful
for understanding the phase transition. In this paper, however, we will
focus on the finite-size case, in which there is no phase transition. 

\section{Main Theorems}

We study the uniqueness of the ground states for two Bose Hubbard models
in the subspace with particle number $N=2n$($n$ is a positive integer).
This work studies finite lattice systems rather than thermodynamic
limits.

For the first model, consider a quantum lattice boson model on an
arbitrary finite-size lattice $\Lambda=\left\{1,2,\dots,L\right\}$ defined by
\begin{align}
H & =H_{0}+H_{U},\\
H_{0} & =-\sum_{i,j\in\Lambda}t_{ij}a_{i}^{+}a_{j},\\
H_{U} & =\sum_{i\in\Lambda}U_{i}\left(a_{i}^{+}a_{i}\right)^{2},
\end{align}
where $a_{i}$ are a set of annihilation operators, and $a_{i}^{+}$
are a set of creation operators. They satisfy the canonical commutation
relations: $\left[a_{s},a_{t}\right]=\left[a_{s}^{+},a_{t}^{+}\right]=0$,
$\left[a_{s},a_{t}^{+}\right]=\delta_{st}$. $t_{ij}=\bar{t}_{ij}$
are real hopping coefficients. We assume $U_{i}<0$ for all $i\in\Lambda$,
which means the interaction is strictly attractive. We further assume
that $\Lambda$  is connected, i.e., starting from any site $i_{1}\in\Lambda$  a particle can hop to
all the sites $i_{k}\in\Lambda (i_{k}\neq i_{1})$ step by step through bonds with $t_{i_{m+1}i_{m}}\neq0$($m=1,\dots,k-1$). Denote the lattice
size as $L=\left|\Lambda\right|$. The particle number $N=\sum_{i\in\Lambda}a_{i}^{+}a_{i}$
is a good quantum number of this model.
We have the following theorem for the first model:

\begin{theorem}\label{thm1}
The ground state in the subspace with particle number
$N=2n$($n$ is a positive integer) for the Bose Hubbard model defined
above is unique.
\end{theorem}

The second model is the two-component Bose Hubbard model. It is a
quantum lattice boson model on an arbitrary finite-size lattice $\Lambda=\left\{1,2,\dots,L\right\}$ defined
by
\begin{align}
H & =H_{0}+H_{U},\\
H_{0} & =-\sum_{i,j\in\Lambda}t_{ij}^{b}b_{i}^{+}b_{j}-\sum_{i,j\in\Lambda}t_{ij}^{c}c_{i}^{+}c_{j},\\
H_{U} & =\sum_{i\in\Lambda}U_{1i}\left(b_{i}^{+}b_{i}+c_{i}^{+}c_{i}\right)^{2}+\sum_{i\in\Lambda}U_{2i}\left(b_{i}^{+}b_{i}-c_{i}^{+}c_{i}\right)^{2},
\end{align}
where $b_{i}$ and $c_{i}$ are two sets of annihilation operators,
and $b_{i}^{+}$ and $c_{i}^{+}$ are two sets of creation operators.
They satisfy the canonical commutation relations: $\left[b_{s},b_{t}\right]=\left[c_{s},c_{t}\right]=\left[b_{s},c_{t}\right]=0$,
$\left[b_{s}^{+},b_{t}^{+}\right]=\left[c_{s}^{+},c_{t}^{+}\right]=\left[b_{s}^{+},c_{t}^{+}\right]=0$,
$\left[b_{s},c_{t}^{+}\right]=\left[b_{s}^{+},c_{t}\right]=0$, $\left[b_{s},b_{t}^{+}\right]=\left[c_{s},c_{t}^{+}\right]=\delta_{st}$.
The complex hopping coefficients of the two components are complex
conjugations of each other, $t_{ij}^{b}=\bar{t}_{ij}^{c}$. For the
interaction terms, we assume that $U_{1i}<0$ and $U_{2i}>0$ for all $i\in\Lambda$.
We also assume that $\Lambda$ is connected,
i.e., starting from any site $i_{1}\in\Lambda$  a particle can hop to
all the sites $i_{k}\in\Lambda (i_{k}\neq i_{1})$ step by step through bonds with $t_{i_{m+1}i_{m}}^{b}\neq0$($m=1,\dots,k-1$).

Define the spin operators on site $i$ by $S_{i}^{\alpha}=\frac{1}{2}\left(\begin{array}{cc}
b_{i}^{+}, & c_{i}^{+}\end{array}\right)\sigma^{\alpha}\left(\begin{array}{c}
b_{i}\\
c_{i}
\end{array}\right)$, and the total spin by $S^{\alpha}=\sum_{i\in\Lambda}S_{i}^{\alpha}$,
where $\sigma^{\alpha}$ are Pauli matrices, $\alpha=x,y,z$.  The particle number
$N=\sum_{i\in\Lambda}\left(b_{i}^{+}b_{i}+c_{i}^{+}c_{i}\right)$
and $S^{z}$ are good quantum numbers of this model.
Obviously this model has spin-$SU\left(2\right)$ invariance when the hopping coefficients $t_{ij}^{b}$ are real.
In this case $\mathbf{S}^{2}$ is also a good quantum number.

We have the following theorem for the second model:

\begin{theorem}\label{thm2}
The ground state in the subspace with particle number
$N=2n$($n$ is a positive integer) for the two-component Bose Hubbard
model defined above is unique. Moreover, for this ground state we have
$S^{z}=0$. When the hopping coefficients are real, it has zero spin quantum
number.
\end{theorem}

\section{Useful Mathematical Facts on Convex Cones}

In this section we introduce some basic knowledge of cone theory which
is used in our proofs, especially the Perron--Frobenius theorem for
cones. For general mathematical accounts of convex cones, the reader
may refer to Ref.~\cite{rockafellar_part_2015,tam_matrices_2006}.

A subset $K$ of a finite-dimensional real vector space $V$ is a
\emph{convex cone} if: (1)for any $\mathbf{x},\mathbf{y}\in K$, we
have $\mathbf{x}+\mathbf{y}\in K$; (2)for any $\mathbf{x}\in K$
and any positive real number $\alpha$ , we have $\alpha\mathbf{x}\in K$.
Moreover, it is said to be a \emph{proper cone} if it satisfies
three further conditions: (1)$K$ is closed in the usual topology
of $V$; (2)$K$ is \emph{pointed}, i.e., if $\mathbf{x},-\mathbf{x}\in K$, we have $\mathbf{x}=\mathbf{0}$;
(3)$\mathrm{int}K\neq\emptyset$, where $\mathrm{int}K$ is the interior
of $K$.

A nonempty set $S$ in a real vector space $V$ \emph{generates} a cone,
denoted $\operatorname{cone}\left(S\right)$,
by $\operatorname{cone}\left(S\right)=\left\{\sum_{i=1}^{m}\lambda_{i}\mathbf{x}_{i}\middle| m\in\mathbb{N},
\lambda_i\geq 0, \mathbf{x}_{i}\in S \right\}$.
The elements of $S$ are called \emph{generators} of $\operatorname{cone}\left(S\right)$.
Define the convex hull of set $S$ by $\operatorname{conv}\left(S\right)=\left\{\sum_{i=1}^{m}\lambda_{i}\mathbf{x}_{i}\middle| m\in\mathbb{N},\lambda_i\geq 0, \mathbf{x}_{i}\in S,\sum_{i}^{m}\lambda_{i}=1 \right\}$.
If $S$ is compact and $\mathbf{0}\notin\operatorname{conv}\left(S\right)$, then $\operatorname{cone}\left(S\right)$
is closed\cite{rockafellar_part_2015}.

For any proper cone $K$ in $\mathbb{R}^{m}$ ($m\geq2$), the vector
$x\in\mathbb{R}^{m}$ is \emph{$K$-positive} if $\mathbf{x}\in\mathrm{int}K$.
The matrix $A\in\mathbb{R}^{m\times m}$ is \emph{$K$-non-negative}
if it maps $K$ to $K$.

Given a closed convex cone $K\subset V$ within the real vector space $V$,
its \emph{dual cone} $K^{*}$ is defined by $K^{*}=\left\{\mathbf{y}\in V\middle|\left\langle \mathbf{y},\mathbf{x}\right\rangle\geq0,
\forall \mathbf{x}\in K \right\}$, where $\left\langle\, , \,\right\rangle$ is the inner product.
A cone is said to be \emph{self-dual} if $K=K^{*}$.

A \emph{face} $F$ of of a cone $K$ is a subset of $K$ which is a cone such that
for any $\mathbf{x}\in F$ and any $\mathbf{y}\in K$ which satisfy
$\mathbf{x}-\mathbf{y}\in K$, we have $\mathbf{y}\in F$.
A face $F$ of $K$ is a \emph{trivial face} if $F=\{\mathbf{0}\}$
or $F=K$. For a subset $S$ of a cone $K$, the intersection of all faces of $K$
including $S$ is called the face of $K$ \emph{generated} by $S$
and is denoted by $\Phi\left(S\right)$(Please notice that $\Phi\left(S\right)$
is not necessarily identical with $\operatorname{cone}\left(S\right).$). A vector $\mathbf{x}\in K$
is an \emph{extreme vector} if either $\mathbf{x}$ is $\mathbf{0}$
or $\mathbf{x}$ is nonzero and $\Phi\left(\{\mathbf{x}\}\right)=\{\lambda\mathbf{x}:\lambda\geq0\}$.
Any face is generated by a set of extreme vectors.

Let $K$ be a closed, pointed convex cone. For any $\mathbf{x}\in K$, the \emph{Carath\'{e}odory number}
of $\mathbf{x}$ is the smallest integer $\kappa\left(\mathbf{x}\right)$
such that $\mathbf{x}=\mathbf{x}_1+\dots+\mathbf{x}_{\kappa\left(\mathbf{x}\right)}$, where
$\mathbf{x}_1,\dots,\mathbf{x}_{\kappa\left(\mathbf{x}\right)}$ belong to extreme rays.
The Carath\'{e}odory number of $K$ is defined as
$\kappa\left(K\right)=\max\left\{\kappa\left(\mathbf{x}\right)\middle|\mathbf{x}\in K\right\}$.
The Carath\'{e}odory theorem gives the bound $\kappa\left(K\right)\leq\dim K$
\cite{rockafellar_part_2015}.

If $A$ is $K$-non-negative, then a face $F$ of $K$ is an \emph{$A$-invariant face}
if $AF\subseteq F$. If $A$ is $K$-non-negative, then $A$ is \emph{$K$-irreducible}
if the only $A$-invariant faces are trivial faces.

We have the following conclusion according to the Perron--Frobenius
theorem for proper cones\cite{tam_matrices_2006}:
\begin{theorem}[Generalization of Perron--Frobenius Theorem]\label{thmPF}
Let $A$ be a $K$-irreducible $K$-non-negative matrix with spectral
radius $\rho$. Then (1)$\rho$ is positive and is a simple eigenvalue
of $A$; (2)there exists a unique $K$-positive eigenvector $\mathbf{u}$
of $A$ corresponding to $\rho$.
\end{theorem}

\section{Proof of Theorem~\ref{thm1}}

Each bosonic many-body wave function with $N$ particles is a full-symmetric
tensor $\Psi_{i_{1},\dots,i_{N}}$, with $\left|\Psi\right\rangle =\sum_{i_{1},\dots,i_{N}}\Psi_{i_{1},\dots,i_{N}}a_{i_{1}}^{+}\cdots a_{i_{N}}^{+}\left|0\right\rangle$, where $i_{1},\dots,i_{N}$ label the sites.
This full-symmetric tensor can be decomposed into a linear combination
of rank-1 full-symmetric tensors(for both complex and real cases)\cite{brachat_symmetric_2010}. The
state corresponding to a rank-1 full-symmetric tensor can
be expressed by $\left|\mathbf{\psi}^{N}\right\rangle =\left(a^{+}\mathbf{\psi}\right)^{N}\left|0\right\rangle $,
where $a^{+}=\left(a_{1}^{+},\dots, a_{L}^{+}\right)$, $\mathbf{\psi}$
is a complex vector in $\mathbb{C}^{L}$. We have $\left|\Psi\right\rangle =\sum_{s}\lambda_{s}\left|\mathbf{\psi}_{s}^{N}\right\rangle $
, where $\lambda_{s}$ are complex numbers. The minimal number of
$\lambda_{s}$ and $\left|\mathbf{\psi}_{s}^{N}\right\rangle $ is
called the rank of $\left|\Psi\right\rangle$, denoted as $\operatorname{rank}\left|\Psi\right\rangle$. The rank of any
state is a finite integer.

For the first model discussed in this work, the Hamiltonian is real
in the occupation-number representation. We may restrict our research
subject to real wave functions $\Psi_{i_{1},\dots,i_{N}}$. In this
case both $\mathbf{\psi}_s$ and $\lambda_{s}$ are real. Furthermore,
$N$=2n is a even positive integer. As shown in Ref.~\cite{purwanto_quantum_2004},
\begin{equation}\label{EqInnerProduct}
\left\langle \mathbf{\psi}_{1}^{2n}\middle|\mathbf{\psi}_{2}^{2n}\right\rangle =\left(2n\right)!\left(\mathbf{\psi}_{1}^{T}\mathbf{\psi}_{2}\right)^{2n}\geq0
\end{equation}
for any two real $\mathbf{\psi}_{1},\mathbf{\psi}_{2}$. When $L>1$, inside this
real subspace $V_{1}$ of the Fock space with $N=2n$ particles, we can define
a proper cone $K_{1}$ by requiring $\lambda_{s}\geq0$ for all $s$.
Obviously any two states in this convex cone have non-negative overlap. We
say the states in $K_{1}$ have Fock space positivity.
From this property we know $K_{1}$ is pointed.

By Eq.~\ref{EqInnerProduct}, we know that $K_{1}\subset K_{1}^{*}$.
We mention that $K_{1}$ is self-dual if $n=1$, and not self-dual if $n>1$. 

The states corresponding to the extreme vectors of $K_{1}$
have the form $\left|\Psi\right\rangle =\lambda\left|\mathbf{\psi}^{2n}\right\rangle$,
where $\lambda\geq0$ and $\mathbf{\psi}\in\mathbb{R}^{L}$. The other
states inside the convex cone $K_{1}$ can be seen as linear combinations
of the states corresponding to the extreme vectors with non-negative coefficients.
$K_{1}$ is generated by the compact set $S=\left\{\left|\mathbf{\psi}^{2n}\right\rangle\middle| \mathbf{\psi}\in S^{L-1}\subset\mathbb{R}^{L}\right\}$, where $S^{L-1}$ is the $\left(L-1\right)$-dimensional
unit sphere centered at the origin of $\mathbb{R}^{L}$,
and $\mathbf{0}\notin \operatorname{conv}\left(S\right)$.
So $K_{1}$ is closed.

For any element $\left|\Psi\right\rangle$ of $K_{1}$, we have
$\left|\Psi\right\rangle=\sum_{i=1}^{\kappa\left(\left|\Psi\right\rangle\right)}\lambda_{i}\left|\mathbf{\psi}_{i}^{2n}\right\rangle$, where $\kappa\left(\left|\Psi\right\rangle\right)$
is the Carath\'{e}odory number of $\left|\Psi\right\rangle$, $\lambda_{i}>0$. By the definition of rank $\kappa\left(\left|\Psi\right\rangle\right)\geq\operatorname{rank}\left|\Psi\right\rangle$.
Let $K^{\prime}$ be the convex cone generated by $\left\{\left|\mathbf{\psi}_{1}^{2n}\right\rangle,\dots,\left|\mathbf{\psi}_{\kappa\left(\left|\Psi\right\rangle\right)}^{2n}\right\rangle\right\}$.
By the Carath\'{e}odory theorem for $K^{\prime}$,
we have $\kappa\left(\left|\Psi\right\rangle\right)\leq\kappa\left(K^{\prime}\right)\leq\operatorname{rank}\left|\Psi\right\rangle$. So $\kappa\left(\left|\Psi\right\rangle\right)=\operatorname{rank}\left|\Psi\right\rangle$.

When $N=2n$, the real subspace $V_{1}$ of the Fock space has dimension $D=\left(\begin{array}{c}
L+2n-1\\
2n
\end{array}\right)$. If an element $\left|\Psi\right\rangle=\sum_{i=1}^{r}\lambda_{i}\left|\mathbf{\psi}_{i}^{2n}\right\rangle$ of $K_{1}$ has rank $r$ less than $D$,
let $W$ be the linear subspace spanned by $\left|\mathbf{\psi}_{1}^{2n}\right\rangle,\dots,\left|\mathbf{\psi}_{r}^{2n}\right\rangle$. Consider $\left|\Psi_{\pm}\right\rangle=\left|\Psi\right\rangle\pm\epsilon\left|\tilde{\Psi}\right\rangle$,
where $\epsilon$ can be any small positive real number, 
$\left|\tilde{\Psi}\right\rangle\in W^{\perp}\subset V_{1}$.
There exists a $\left|\Psi^{\prime}\right\rangle\in K_{1}^{*}\backslash W$
such that $\left\langle\Psi^{\prime}\middle|\Psi_{\pm}\right\rangle\neq 0$,
and at least one of them is negative.
So $\left|\Psi\right\rangle$ must belong to the boundary of $K_{1}$.
If an element $\left|\Psi\right\rangle$ of $K_{1}$ has rank $D$,
it remains in $K_{1}$ after any infinitesimal change in $V_{1}$,
since we can set all the coefficients $\lambda_{i}>0$, $i=1,\dots,D$.
So any element of $K_{1}$ with rank $D$ belongs to its interior.

For any element $\left|\Psi\right\rangle$ in $K_{1}$, we can define
the kernel of $\left|\Psi\right\rangle$ by
\begin{equation}
\ker\left|\Psi\right\rangle=\left\{\mathbb{\mathbf{\psi}}\in\mathbb{R}^{L}\middle|\left\langle\mathbf{\psi}^{2n}\middle|\Psi\right\rangle=0\right\} .
\end{equation}
Clearly, by Eq.~\ref{EqInnerProduct}, if $\mathbf{\psi}_{1},\mathbf{\psi}_{2}\in\ker\left|\Psi\right\rangle$
and $r_{1},r_{2}\in\mathbb{R}$, then $r_{1}\mathbf{\psi}_{1}+r_{2}\mathbf{\psi}_{2}\in\ker\left|\Psi\right\rangle$.
So $\ker\left|\Psi\right\rangle$ is a linear subspace of $\mathbb{R}^{L}$.
The range of $\left|\Psi\right\rangle $ is defined by $\operatorname{range}\left|\Psi\right\rangle=\left(\ker\left|\Psi\right\rangle\right)^{\perp}$.
Define the linear rank of $\left|\Psi\right\rangle$ as 
$\operatorname{linearrank}\left|\Psi\right\rangle=\dim\operatorname{range}\left|\Psi\right\rangle$.
Please notice that the definition of "linear rank" is different from the definition of "rank".

Given any subspace $L_{0}$ of $\mathbb{R}^{L}$, we can define a
subset $F_{1,L_{0}}$ of $K_{1}$ by
\begin{equation}
F_{1,L_{0}}=\left\{ \left|\Psi\right\rangle \in K_{1}\middle|\operatorname{range}\left|\Psi\right\rangle \subset L_{0}\right\} .
\end{equation}
For any element $\left|\Psi\right\rangle \in F_{1,L_{0}}$, suppose
$\left|\Psi\right\rangle =\left|\Psi_{1}\right\rangle +\left|\Psi_{2}\right\rangle $, and
$\left|\Psi_{1}\right\rangle ,\left|\Psi_{2}\right\rangle \in K_{1}$.
If $\mathbf{\psi}\in\ker\left|\Psi\right\rangle $, then we have $\mathbf{\psi}\in\ker\left|\Psi_{1}\right\rangle $
and $\mathbf{\psi}\in\ker\left|\Psi_{2}\right\rangle $. So $\operatorname{range}\left|\Psi_{1}\right\rangle \subset\operatorname{range}\left|\Psi\right\rangle \subset L_{0}$, and
$\operatorname{range}\left|\Psi_{2}\right\rangle \subset\operatorname{range}\left|\Psi\right\rangle \subset L_{0}$.
Hence $\left|\Psi_{1}\right\rangle ,\left|\Psi_{2}\right\rangle \in F_{1,L_{0}}$
and $F_{1,L_{0}}$ is indeed a face of $K_{1}$.

Suppose $F_{1}$ is a non-empty face of $K_{1}$. Select an element
$\left|\Psi\right\rangle \in F_{1}$ so that $\left|\Psi\right\rangle $
has the largest rank(thus the largest Carath\'{e}odory number and linear rank) among all the elements in $F_{1}$. Let $L_{0}=\operatorname{range}\left|\Psi\right\rangle $.
Now we show that $F_{1}=F_{1,L_{0}}$. Assume $\left|\Psi^{\prime}\right\rangle \in F_{1}$
but $\left|\Psi^{\prime}\right\rangle \notin F_{1,L_{0}}$. Then there
exits $\mathbb{\mathbf{\psi}}$ such that $\left\langle \mathbf{\psi}^{2n}\middle|\Psi\right\rangle =0$
but $\left\langle \mathbf{\psi}^{2n}\middle|\Psi^{\prime}\right\rangle >0$.
For $\left|\Psi\right\rangle +\left|\Psi^{\prime}\right\rangle \in F_{1}$,
we have $\operatorname{linearrank}\left(\left|\Psi\right\rangle +\left|\Psi^{\prime}\right\rangle \right)>\operatorname{linearrank}\left(\left|\Psi\right\rangle \right)$,
which is a contradiction. So we must have $F_{1}\subset F_{1,L_{0}}$.
Since $\left|\Psi\right\rangle $ is in the interior of $F_{1,L_{0}}$,
for any $\left|\tilde{\Psi}\right\rangle \in F_{1,L_{0}}$ there exists
a sufficient small positive number $\epsilon$ such that $\left|\Psi\right\rangle -\epsilon\left|\tilde{\Psi}\right\rangle \in F_{1,L_{0}}$.
So $\left|\tilde{\Psi}\right\rangle $ also belongs to $F_{1}$. Hence
we have $F_{1,L_{0}}\subset F_{1}$. Combining the two sides is $F_{1}=F_{1,L_{0}}$.

For any positive real $\beta$, the eigenvector corresponding to the
spectral radius of $\exp\left(-\beta H\right)$ is exactly the ground
state. Let $\beta=M\tau$($M$ is a positive integer) and carry out
the following Trotter-Suzuki decomposition:
\begin{equation}
\exp\left(-\beta H\right)=\left[\exp\left(-\tau H_{0}\right)\exp\left(-\tau H_{U}\right)\right]^{M}+O\left(M\tau^{2}\right).
\end{equation}
The error term scales as $O\left(M\tau^{2}\right)$ and will disappear
as $M$ approaches infinity. Then for each $\exp\left(-\tau H_{U}\right)$
and each site, we carry out the Hubbard-Stratonovich transformation:
\begin{align}
 & \exp\left[-\tau U_{i}\left(a_{i}^{+}a_{i}\right)^{2}\right]\nonumber \\
 & =\sqrt{-\frac{1}{4\pi\tau U_{i}}}\int_{-\infty}^{+\infty}\exp\left[\frac{x_{i}^{2}}{4\tau U_{i}}-\left(a_{i}^{+}a_{i}\right)x_{i}\right]dx_{i}.
\end{align}
Thus the exponential of the quartic forms of creation and annihilation
operators can be expressed as the Gaussian integral of the exponential
of some quadratic forms. For the exponential of the quadratic forms, it is
easy to show that $\exp\left(a^{+}Ta\right)\left|\mathbf{\mathbf{\psi}}^{N}\right\rangle =\left|\left[\exp\left(T\right)\mathbf{\mathbf{\psi}}\right]^{N}\right\rangle $
for any square matrix $T$. Thus by acting $\exp\left(-\beta H\right)$
on any element of $K_{1}$, we have shown that $\exp\left(-\beta H\right)$
is a $K_{1}$-non-negative matrix.

According to Theorem~\ref{thmPF}, to show that the ground state of $H$ in the subspace with $N=2n$
particles is unique, it is sufficient to show that $\exp\left(-\beta H\right)$ is a  $K_{1}$-irreducilble $K_{1}$-non-negative matrix
in this subspace. This can be done if we can prove that for any two non-zero extreme vectors $\left|\mathbf{\psi}^{N}_{0}\right\rangle$
and $\left|\mathbf{\psi}^{N}_{1}\right\rangle$,
we have $\left\langle \mathbf{\psi}^{N}_{1}\right|e^{-\beta H}\left|\mathbf{\psi}^{N}_{0}\right\rangle >0$.

Due to the Trotter-Suzuki decomposition and the Hubbard-Stratonovich transformation, the action of $\exp\left(-t H\right)$($t\in\left[0,\beta\right]$) on any non-zero
extreme vector $\left|\mathbf{\psi}^{N}_{0}\right\rangle $ in $K_{1}$
can be seen as an It\^o process on $\mathbb{R}^{L}$. It gives a
parameterized collection of random variables $\left\{ \mathbf{\Psi}_{t}\right\} _{t\in\left[0,\beta\right]}$
on $\mathbb{R}^{L}$, which satisfy the following stochastic differential
equation:
\begin{equation}
d\mathbf{\Psi}_{t}=-T_{0}\mathbf{\Psi}_{t}dt-\sum_{i\in\Lambda}\sqrt{2\left|U_{i}\right|}T_{i}\mathbf{\Psi}_{t}dB^{i}_{t},\ \mathbf{\Psi}_{0}=\psi_{0}.
\end{equation}
where the matrix $T_{0}$ satisfies $a^{+}T_{0}a=H_{0}$, $T_{i}$
satisfies $a^{+}T_{i}a=a^{+}_{i}a_{i}$, $\mathbf{B}_{t}=\left(B^{1}_{t},B^{2}_{t},\dots,B^{L}_{t}\right)^{T}$
is the $L$-dimensional standard Brownian motion. We can rewrite this
It\^o form of stochastic differential equation into Stratonovich
form\cite{oksendal_ito_2003}:
\begin{equation}
d\mathbf{\Psi}_{t}=V_{0}\left(\mathbf{\Psi}_{t}\right)dt+\sum_{i\in\Lambda}V_{i}\left(\mathbf{\Psi}_{t}\right)\circ dB^{i}_{t},\ \mathbf{\Psi}_{0}=\psi_{0},
\end{equation}
where $V_{0}\left(\mathbf{\Psi}_{t}\right)=\left(-T_{0}+\sum_{i\in\Lambda}U_{i}T_{i}^{2}\right)\mathbf{\Psi}_{t}$,
$V_{i}\left(\mathbf{\Psi}_{t}\right)=-\sqrt{2\left|U_{i}\right|}T_{i}\mathbf{\Psi}_{t}$.

According to the corollary of the Stroock--Varadhan support theorem\cite{victoir_support_2010}, the solutions to the ordinary differential equation 
\begin{equation}
d\mathbf{\psi}=V_{0}\left(\mathbf{\psi}\right)dt+\sum_{i\in\Lambda}V_{i}\left(\mathbf{\psi}\right)\circ dh^{i},\ \mathbf{\psi}\left(t=0\right)=\mathbf{\psi}_{0}
\end{equation}
decide the topological support of $\mathbf{\Psi}$, where $\mathbf{h}\in W^{1,2}_{0}\left(\left[0,\beta\right],\mathbb{R}^{L}\right)$.
By studying the Lie semigroup generated by matrices $-T_{0}+\sum_{i\in\Lambda}U_{i}T_{i}^{2}$
and $\pm\sqrt{2\left|U_{i}\right|}T_{i}$ , one can show that $\psi\left(\beta\right)$
can reach almost any point in at least one $L$-dimensional orthant
as $\mathbf{h}$ changes. The topological support of $\mathbf{\Psi}_{\beta}$
is contains at least one $L$-dimensional orthant, and thus
\begin{equation}
\left\langle \mathbf{\psi}^{N}_{1}\right|e^{-\beta H}\left|\mathbf{\psi}^{N}_{0}\right\rangle =\mathbb{E}\left(\left\langle \mathbf{\psi}^{N}_{1}\middle|\mathbf{\Psi}^{N}_{\beta}\right\rangle \right)>0
\end{equation}
for any two non-zero extreme vectors $\left|\mathbf{\psi}^{N}_{0}\right\rangle $
and $\left|\mathbf{\psi}^{N}_{1}\right\rangle$. Q.E.D.

\section{Proof of Theorem~\ref{thm2}}

The proof of Theorem~\ref{thm2} is very similar to that of Theorem~\ref{thm1}, except
some subtle differences listed below.

In this case the state corresponding to a rank-1 full-symmetric
tensor can be expressed by $\left|\mathbf{\psi}^{N}\right\rangle =\left(a^{+}\mathbf{\psi}\right)^{N}\left|0\right\rangle $,
where $a^{+}=\left(b_{1}^{+},\dots, b_{L}^{+},c_{1}^{+},\dots, c_{L}^{+}\right)$,
$\mathbf{\psi}$ is a complex vector in $\mathbb{C}^{2L}$. For any
arbitrary state $\left|\Psi\right\rangle $ in this subspace of the Fock space with
$N$ particles, we still have $\left|\Psi\right\rangle =\sum_{s}\lambda_{s}\left|\mathbf{\psi}_{s}^{N}\right\rangle $
, where $\lambda_{s}$ are complex numbers, $\psi_s\in\mathbb{C}^{2L}$.

We may restrict our research subject to a real subspace of the subspace
of the Fock space with $N=2n$ particles. Consider a special kind
of $\mathbf{\psi}$: $\mathbf{\psi}=\left(\begin{array}{c}
\mathbf{\phi}\\
\bar{\mathbf{\phi}}
\end{array}\right)$, $\phi\in\mathbb{C}^{L}$. This kind of $\mathbf{\psi}$ span a real
subspace of $\mathbb{C}^{2L}$, denoted as $\tilde{\mathbf{\mathbb{R}}}^{2L}$.
We still have
\begin{equation}
\left\langle \mathbf{\psi}_{1}^{2n}\middle|\mathbf{\psi}_{2}^{2n}\right\rangle =\left(2n\right)!\left(\mathbf{\psi}_{1}^{T}\mathbf{\psi}_{2}\right)^{2n}\geq0
\end{equation}
for any $\mathbf{\psi}_{1},\mathbf{\psi}_{2}\in\tilde{\mathbf{\mathbb{R}}}^{2L}$.
Hence this kind of states $\left|\mathbf{\psi}^{2n}\right\rangle $($\mathbf{\psi}\in\tilde{\mathbf{\mathbb{R}}}^{2L}$)
span a real subspace $V_{2}$ of the subspace of the Fock space with $N=2n$
particles(In fact the subspace of the Fock space with $N=2n$ particles
can be seen as a complexification of this real subspace.). As we will
show below, the Hamiltonian $H$ of the second model is a real matrix
on this real subspace. Inside this real subspace $V_{2}$, we define a
proper cone $K_{2}$ by requiring $\lambda_{s}\geq0$, $\mathbf{\psi}_{s}\in\tilde{\mathbf{\mathbb{R}}}^{2L}$
for all $s$. Any two wave functions in this cone again have non-negative
overlap. We say the states in $K_{2}$ have Fock space reflection
positivity. It's easy to show that $K_{2}$ is closed and pointed. $K_{2}\subset K_{2}^{*}$.

The states corresponding to the extreme vectors of $K_{2}$ have the form $\left|\Psi\right\rangle =\lambda\left|\mathbf{\psi}^{2n}\right\rangle $,
where $\lambda\geq0$ and $\mathbf{\psi}\in\tilde{\mathbf{\mathbb{R}}}^{2L}$.
The other states inside the convex cone $K_{2}$ can be seen as linear combinations
of the states corresponding to the extreme vectors with non-negative coefficients.

When $N=2n$, the real subspace $V_{2}$ of the Fock space has dimension $D=\left(\begin{array}{c}
2L+2n-1\\
2n
\end{array}\right)$. All the elements in the interior of $K_{2}$ has rank $D$,
and any element of $K_{2}$ with rank $D$ belongs to its interior.

For any element $\left|\Psi\right\rangle $ in $K_{2}$, we can define
the kernel of $\left|\Psi\right\rangle $ by
\begin{equation}
\ker\left|\Psi\right\rangle=\left\{ \mathbb{\mathbf{\psi}}\in\tilde{\mathbb{R}}^{2L}\middle|\left\langle \mathbf{\psi}^{2n}\middle|\Psi\right\rangle=0\right\} .
\end{equation}
Clearly, if $\mathbf{\psi}_{1},\mathbf{\psi}_{2}\in\ker\left|\Psi\right\rangle $
and $r_{1},r_{2}\in\mathbb{R}$, then $r_{1}\mathbf{\psi}_{1}+r_{2}\mathbf{\psi}_{2}\in\ker\left|\Psi\right\rangle $.
So $\ker\left|\Psi\right\rangle $ is a linear subspace of $\tilde{\mathbb{R}}^{2L}$.
The range of $\left|\Psi\right\rangle $ is defined by $\operatorname{range}\left|\Psi\right\rangle =\left(\ker\left|\Psi\right\rangle \right)^{\perp}$. Any face of $K_{2}$ can be written as
\[
F_{2,L_{0}}=\left\{ \left|\Psi\right\rangle \in K_{2}\middle|\operatorname{range}\left|\Psi\right\rangle \subset L_{0}\right\} ,
\]
for any subspace $L_{0}$ of $\tilde{\mathbb{R}}^{2L}$. The proof
is very similar to that of $K_{1}$, omitted.

We carry out the Trotter-Suzuki decomposition and the Hubbard-Stratonovich
transformation for $\exp\left(-\beta H\right)$, for any positive real
$\beta$. There are two types of Hubbard-Stratonovich transformations
in this case:
\begin{align}
 & \exp\left[-\tau U_{1i}\left(b_{i}^{+}b_{i}+c_{i}^{+}c_{i}\right)^{2}\right]\nonumber \\
 & =\sqrt{-\frac{1}{4\pi\tau U_{1i}}}\int_{-\infty}^{+\infty}\exp\left[\frac{x_{1i}^{2}}{4\tau U_{1i}}-\left(b_{i}^{+}b_{i}+c_{i}^{+}c_{i}\right)x_{1i}\right]dx_{1i},
\end{align}
\begin{align}
 & \exp\left[-\tau U_{2i}\left(b_{i}^{+}b_{i}-c_{i}^{+}c_{i}\right)^{2}\right]\nonumber \\
 & =\sqrt{\frac{1}{4\pi\tau U_{2i}}}\int_{-\infty}^{+\infty}\exp\left[-\frac{x_{2i}^{2}}{4\tau U_{2i}}-i\left(b_{i}^{+}b_{i}-c_{i}^{+}c_{i}\right)x_{2i}\right]dx_{2i}.
\end{align}
Obviously the action of the exponential of the quadratic forms here on
$\left|\mathbf{\psi}^{2n}\right\rangle $ keeps the vector $\mathbf{\psi}$
in $\tilde{\mathbf{\mathbb{R}}}^{2L}$. Hence $\exp\left(-\beta H\right)$
is a $K_{2}$-non-negative matrix. So $1-d\tau H$ should be a real
matrix on this real vector space, so does $H$.

To show that the ground state of $H$ in the subspace with $N=2n$
particles is unique, it is sufficient to show that $\exp\left(-\beta H\right)$ is
a $K_{2}$-irreducible $K_{2}$-non-negative matrix in this subspace.
This can be done if we can prove that for any two non-zero extreme vectors $\left|\mathbf{\psi}^{N}_{0}\right\rangle$
and $\left|\mathbf{\psi}^{N}_{1}\right\rangle$,
we have $\left\langle \mathbf{\psi}^{N}_{1}\right|e^{-\beta H}\left|\mathbf{\psi}^{N}_{0}\right\rangle >0$.

The action of $\exp\left(-t H\right)$($t\in\left[0,\beta\right]$) on any non-zero
extreme vector $\left|\mathbf{\psi}^{N}_{0}\right\rangle $ in $K_{2}$
can be seen as an It\^o process on $\tilde{\mathbb{R}}^{2L}$. It
gives a parameterized collection of random variables $\left\{ \mathbf{\Psi}_{t}\right\} _{t\in\left[0,\beta\right]}$
on $\tilde{\mathbb{R}}^{2L}$, which satisfy the following stochastic
differential equation:
\begin{equation}
d\mathbf{\Psi}_{t}=-T_{0}\mathbf{\Psi}_{t}dt-\sum_{i\in\Lambda}\sqrt{2\left|U_{1i}\right|}T_{1i}\mathbf{\Psi}_{t}dB^{i}_{t}-\sum_{i\in\Lambda}\sqrt{2U_{2i}}T_{2i}\mathbf{\Psi}_{t}dB^{i+L}_{t},\ \mathbf{\Psi}_{0}=\psi_{0}.
\end{equation}
where the matrix $T_{0}$ satisfies $a^{+}T_{0}a=H_{0}$, $T_{1i}$
satisfies $a^{+}T_{1i}a=b^{+}_{i}b_{i}+c^{+}_{i}c_{i}$, $T_{2i}$
satisfies $a^{+}T_{2i}a=i\left(b^{+}_{i}b_{i}-c^{+}_{i}c_{i}\right)$,
$\mathbf{B}_{t}=\left(B^{1}_{t},B^{2}_{t},\dots,B^{2L}_{t}\right)^{T}$
is the $2L$-dimensional standard Brownian motion. We can rewrite
this It\^o form of stochastic differential equation into Stratonovich
form:
\begin{equation}
d\mathbf{\Psi}_{t}=V_{0}\left(\mathbf{\Psi}_{t}\right)dt+\sum_{i\in\Lambda}V_{1i}\left(\mathbf{\Psi}_{t}\right)\circ dB^{i}_{t}+\sum_{i\in\Lambda}V_{2i}\left(\mathbf{\Psi}_{t}\right)\circ dB^{i+L}_{t},\ \mathbf{\Psi}_{0}=\psi_{0},
\end{equation}
where $V_{0}\left(\mathbf{\Psi}_{t}\right)=\left(-T_{0}+\sum_{i\in\Lambda}U_{1i}T_{1i}^{2}-\sum_{i\in\Lambda}U_{2i}T_{2i}^{2}\right)\mathbf{\Psi}_{t}$,
$V_{1i}\left(\mathbf{\Psi}_{t}\right)=-\sqrt{2\left|U_{1i}\right|}T_{1i}\mathbf{\Psi}_{t}$,
$V_{2i}\left(\mathbf{\Psi}_{t}\right)=-\sqrt{2U_{2i}}T_{2i}\mathbf{\Psi}_{t}$.

According to the corollary of the Stroock--Varadhan support theorem,
the solutions to the ordinary differential equation 
\begin{equation}
d\mathbf{\psi}=V_{0}\left(\mathbf{\psi}\right)dt+\sum_{i\in\Lambda}V_{1i}\left(\mathbf{\psi}\right)dh^{i}++\sum_{i\in\Lambda}V_{2i}\left(\mathbf{\psi}\right)dh^{i+L},\ \mathbf{\psi}\left(t=0\right)=\mathbf{\psi}_{0}
\end{equation}
decide the topological support of $\mathbf{\Psi}$, where $\mathbf{h}\in W^{1,2}_{0}\left(\left[0,\beta\right],\mathbb{R}^{2L}\right)$.
By studying the Lie semigroup generated by matrices $-T_{0}+\sum_{i\in\Lambda}U_{1i}T_{1i}^{2}-\sum_{i\in\Lambda}U_{2i}T_{2i}^{2}$,
$\pm\sqrt{2\left|U_{1i}\right|}T_{1i}$ and $\pm\sqrt{2U_{2i}}T_{2i}$,
one can show that $\psi\left(\beta\right)$ can reach almost any point
in $\tilde{\mathbb{R}}^{2L}\backslash\left\{ \mathbf{0}\right\} $
as $\mathbf{h}$ changes. The topological support of $\mathbf{\Psi}_{\beta}$
is $\tilde{\mathbb{R}}^{2L}$, and thus
\begin{equation}
\left\langle \mathbf{\psi}^{N}_{1}\right|e^{-\beta H}\left|\mathbf{\psi}^{N}_{0}\right\rangle =\mathbb{E}\left(\left\langle \mathbf{\psi}^{N}_{1}\middle|\mathbf{\Psi}^{N}_{\beta}\right\rangle \right)>0
\end{equation}
for any two non-zero extreme vectors $\left|\mathbf{\psi}^{N}_{0}\right\rangle$
and $\left|\mathbf{\psi}^{N}_{1}\right\rangle$.

The ground state of $H$ in the subspace with $N=2n$ particles must
has Fock space reflection positivity according to the Perron--Frobenius
theorem for cones. Consider the overlaps between $\left|\mathbf{\psi}^{2n}\right\rangle $($\mathbf{\psi}\in\tilde{\mathbf{\mathbb{R}}}^{2L}$,
$\mathbf{\psi}$ is nonzero) and the following states with $S^{z}=0$:
$\left(b_{_{1}}^{+}\right)^{i_{1}}\left(c_{_{1}}^{+}\right)^{i_{1}}\dots\left(b_{L}^{+}\right)^{i_{L}}\left(c_{L}^{+}\right)^{i_{L}}\left|0\right\rangle $($i_{1}+\dots+i_{L}=n$).
A direct calculation shows that there exists at least one such singlet
state which has positive overlap with $\left|\mathbf{\psi}^{2n}\right\rangle$.
Therefore the ground state must have $S^{z}=0$. When the hopping coefficients are real,
i.e., when there is spin-$SU\left(2\right)$ invariance,
the unique ground state must have zero spin quantum number. Q.E.D.

\section{Concluding Remarks}

We have prooved the uniqueness of the ground
states for two Bose Hubbard models with particle number $N=2n$($n$
is a positive integer). Our proofs are independent of lattice shapes and dimensions.

In our theorems, the ground state always
belongs to the interior of the related convex cone(and thus the interior of its dual cone).
This means the ground state always have strictly positive overlap with
any non-zero extreme vector $\left|\mathbf{\psi}^{2n}\right\rangle$.
The two positivities discussed in our work lead to this nodeless
feature of the ground state wave function.

If we change the conditions for the first Bose Hubbard model to:
(1)$t_{ij}\geq0$ for all $i,j\in\Lambda$; (2)$\Lambda$  is connected,
i.e., starting from any site $i_{1}\in\Lambda$  a particle can hop to
all the sites $i_{k}\in\Lambda (i_{k}\neq i_{1})$ step by step through bonds with $t_{i_{m+1}i_{m}}\neq0$($m=1,\dots,k-1$);
(3)$U_i$ can take arbitrary real number for all $i\in\Lambda$;
then $\exp(-\beta H)$($\beta$ is a positive real number)
is an irreducible non-negative matrix in the occupation-number representation.
We can prove the uniqueness of the ground states for this model
for both even and odd particle numbers $N$, using the Perron--Frobenius theorem
for matrices with non-negatice elements.

It is not only important to study these two Bose Hubbard models from an analytical
perspective, but also from a numerical perspective. There is a deep
connection between positivities and the absence of the sign problem
in quantum Monte Carlo simulations, as pointed out in Ref.~\cite{wei_majorana_2016}.
Since any two states with Fock space positivity have non-negative
overlap, the first model discussed in this work can be simulated using
projector quantum Monte Carlo without encountering any sign problem\cite{purwanto_quantum_2004}.
We would like to point out that a similar algorithm should also work
for the second model, for any two states with Fock space reflection
positivity also have non-negative overlap.

In Ref.~\cite{wei_semigroup_2024} two types of quantum lattice
fermion models without the sign problem in quantum Monte Carlo simulations were proposed.
For one type of those models this sign-problem-free property can be explained by
Majorana reflection positivity, while the explanation for the other
type is still missing. We point out here that a new positivity,
which we may call time-reversal positivity, has been implied in that
work already. Exploring the connection between the uniqueness of the
ground states for time-reversal invariant quantum lattice fermion
models and this new time-reversal positivity could be very interesting.
We will keep it for further research work.

\backmatter

\bmhead{Acknowledgements}
We would like to thank Yin-Kai Yu for helpful discussions on quantum
Monte Carlo simulations. C. Z. is supported by the National Natural
Science Foundation of China(No. 12071253).

\section*{Declarations}
\begin{itemize}
\item Funding

This work is supported by the National Natural Science Foundation of China(No. 12071253).
\item Competing interests

The authors have no relevant financial or non-financial interests to disclose.
\item Ethics approval and consent to participate

Not applicable.
\item Consent for publication

All authors have read and agreed.
\item Data availability

Not applicable.
\item Materials availability

Not applicable.
\item Code availability

Not applicable.
\item Author contribution

\textbf{Zhong-Chao Wei}: Conceptualization (lead); writing -- original draft (lead); formal analysis (lead); writing -- review and editing (equal).
\textbf{Chong Zhao}: Methodology (lead); writing -- review and editing (equal).
\end{itemize}

\bibliography{BoseHubbardLieb}

\end{document}